\documentclass{article}
\usepackage{graphicx} % Required for inserting images
\usepackage[english]{babel}
\usepackage[square,numbers]{natbib}
\title{A Framework for LLM-powered Design Assistants}
\author{Swaroop Panda \\ \small\textbf{Northumbria University}}
\date{}

\begin{document}

\maketitle
\begin{abstract}
    Design assistants are frameworks, tools or applications intended to facilitate both the creative and technical facets of design processes. Large language models (LLMs) are AI systems engineered to analyze and produce text resembling human language, leveraging extensive datasets. This study introduces a framework wherein LLMs are employed as Design Assistants, focusing on three key modalities within the Design Process: Idea Exploration, Dialogue with Designers, and Design Evaluation. Importantly, our framework is not confined to a singular design process but is adaptable across various processes. 
\end{abstract}


\section{Introduction}
Design assistants, leveraging advancements in artificial intelligence and machine learning, have become indispensable tools in various creative and technical fields \cite{kim2019ai,lee2020guicomp,huet2021context}. These systems, often integrated into software applications, provide designers with real-time suggestions, automate repetitive tasks, and facilitate complex problem-solving processes. By analyzing vast datasets and recognizing patterns, design assistants can propose optimized solutions \cite{koyama2022bo}, enhance productivity \cite{huet2021context}, and foster innovation. Their capabilities extend to diverse domains such as graphic design, architecture, and product development, where they assist in generating novel ideas, ensuring adherence to design principles, and maintaining consistency. The incorporation of design assistants into the creative workflow not only augments human creativity but also bridges the gap between conceptualization and execution, ultimately contributing to more efficient and effective design processes \cite{lee2020guicomp,huet2021context}.

The popularity of Large Language Models (LLMs) has surged in recent years, driven by their exceptional performance in a wide array of natural language processing tasks \cite{zhao2023survey}. This ascent is attributable to several factors, including advancements in computational power, the availability of vast datasets for training, and innovations in model architectures such as transformers. LLMs have demonstrated remarkable capabilities in generating coherent and contextually relevant text, understanding and responding to queries, and performing complex linguistic tasks with minimal supervision \cite{li2024pre}. Their utility spans diverse applications, from automated content creation and customer service to sophisticated academic research and data analysis, underscoring their transformative impact on both industry and academia. Consequently, the research community has witnessed a proliferation of studies aimed at refining these models, addressing their ethical implications, and exploring their potential to advance artificial intelligence further \cite{stella2023can,duan2023towards}.

In this paper, we propose a framework outlining the role of LLMs as Design Assistants. Our framework is not conditioned on any specific design process, rather, it encompasses modalities that can be partially or fully integrated into various design processes.  The modalities included in our framework are Idea Exploration, Crafting Dialogues with Designers, and Design Evaluation.

% The contribution of this paper are three fold,
% \begin{enumerate}
%     \item We present a framework of how LLMs can act as Design Assistants; derived from the capabilities of LLMs and existing Design Research literature.
%     \item 
%     \item We provide examples of exemplar prompts for LLMs using different design case studies, within the framework.
%     % \item We analyse responses from LLMs and then suggest and critique as assistants in the design process.
% \end{enumerate}

\section{Background}
\subsection{Design Processes \& Assistants}
Design processes have been a focal point of research within the fields of engineering, architecture, and product development, emphasizing the intricate balance between creativity and systematic problem-solving. Simon \cite{simon1969sciences} introduced the concept of design as a form of problem-solving, highlighting the iterative nature of the design process wherein designers continually redefine problems and solutions. Cross \cite{cross1982designerly} further elaborated on this, proposing that design thinking encompasses distinct cognitive strategies that differentiate it from other problem-solving methods. In exploring these strategies, Dorst and Cross \cite{dorst2001creativity} identified the "co-evolution of problem and solution spaces" as a critical characteristic of effective design processes, where designers concurrently develop a deeper understanding of the problem while exploring potential solutions. Lawson \cite{lawson2005designers} emphasized the importance of both divergent and convergent thinking in the design process, arguing that successful design outcomes often emerge from a dynamic interplay between these modes of thought. This body of work collectively underscores the complexity and multifaceted nature of design processes, illustrating how iterative cycles of ideation, evaluation, and refinement are essential to achieving innovative and effective design solutions.

\subsection{Why use LLMs for Design Assistance}
Beyond concept generation \cite{barua2024concept} and feedback, LLMs enhance collaborative design environments and assist with background research \cite{aubin2024llms}, cultural sensitivity analysis \cite{adilazuarda2024towards}, and content generation. In collaborative tools, LLMs facilitate communication by summarizing discussions, tracking design decisions, and suggesting compromises. They conduct background research \cite{whitfield2023elicit}, summarizing literature and cultural contexts \cite{jin2024comprehensive} relevant to the project. Further, LLMs are trained on large datasets. LLMs such as those based on the GPT architecture \cite{openai_chatgpt}, derive their efficacy from extensive training on large datasets, encompassing diverse linguistic and contextual information. This robust training equips LLMs with the capability to contribute significantly to various stages of design processes. Their proficiency lies in their ability to generate coherent and contextually appropriate text, provide insights through language-based analyses \cite{zhang2023redefining}, and even assist in ideation and conceptualization phases by offering creative prompts or refining conceptual frameworks. Moreover, LLMs can enhance user interaction and experience design by generating natural language responses that simulate human-like communication \cite{cai2023does}, thereby broadening their utility in applications ranging from user interface design to content creation and beyond.

\section{Anatomy of the Framework}
The proposed framework is structured around various checkpoints where LLMs can offer support to designers throughout the design process. Crucially, the framework is not intended to replace the designer, nor to direct or control the design process. Instead, its purpose is to assist and enhance the designer's capabilities, providing supplementary insights and tools that can be utilized as deemed appropriate by the designer. This approach ensures that the designer retains full control over the creative process while benefiting from the augmented capabilities provided by AI. The proposed framework is intentionally designed to be adaptable, devoid of any presupposition regarding a specific design process. It is structured to integrate seamlessly into various design methodologies, whether partially or fully. This flexibility ensures that the framework can accommodate diverse design practices and can be tailored to meet the unique requirements of different projects and organizational contexts. 

% Consequently, it provides a versatile tool that enhances the robustness and applicability of the design process across various domains and stages.
\section{The Framework}
The framework consists of three different modalities; Idea Exploration, Crafting Dialogue with Designers and Design Evaluation.

\begin{figure*}[!htb]
    \centering
    \framebox{
    \includegraphics[width=\textwidth]{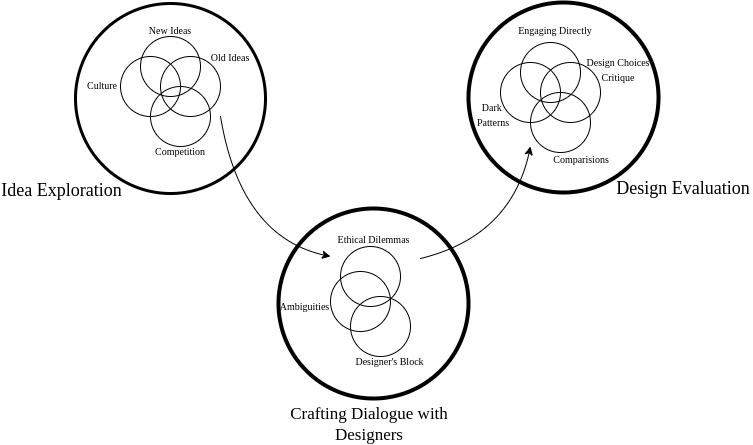}}
    \caption{The framework with three different modalities. Figure adapted from (\cite{takeda1990modeling})}
    \label{fig:enter-label3}
\end{figure*}

\subsection{Idea Exploration}
% Idea exploration is an important part of the design process. This stage not only facilitates the generation of innovative concepts but also serves to refine and validate potential solutions through iterative analysis and creative deliberation.
Idea exploration constitutes a pivotal phase within the design process, wherein diverse concepts are generated and assessed. This stage is instrumental in fostering innovation, as it encourages creative thinking and the examination of multiple perspectives. Furthermore, idea exploration enables designers to refine and validate potential solutions through iterative analysis. This iterative approach facilitates continuous feedback and creative deliberation, allowing for the identification of strengths and weaknesses in proposed ideas. Ultimately, engaging in thorough idea exploration enhances the likelihood of developing effective and user-centered solutions, thereby contributing to the overall success of the design endeavor.

\begin{enumerate}
    \item \textbf{New Ideas} LLMs serve as pivotal tools in generating innovative ideas. They employ vast datasets to comprehend and synthesize human language and thus different ideas. By harnessing this capability, LLMs can explore diverse sources of inspiration, ranging from literature, non-fiction and technical manuals to contemporary discourse on design theory. Additionally, through fine-tuned parameters and tailored prompts , LLMs can be directed to engage in creative brainstorming, producing novel concepts that may elude traditional design approaches \cite{shaer2024ai}. Furthermore, their ability to adapt and iterate based on feedback enhances the iterative design process, fostering the evolution of ideas into tangible innovations \cite{gramopadhye2024few}. 
    
    % So LLMs offer a sophisticated avenue for designers to access and cultivate new design paradigms, leveraging the amalgamation of computational prowess and linguistic nuance to transcend conventional boundaries in design thinking.
    \item \textbf{Cultural Sensitivities}: LLMs possess the capability to comprehend and generate language across multiple cultural contexts, thereby facilitating nuanced discussions on cultural diversity in design \cite{adilazuarda2024towards}. By incorporating diverse linguistic datasets and training models on a broad spectrum of cultural expressions, LLMs can identify and respect cultural nuances in design elements such as color symbolism, iconography, textural preferences among others. Thus they can contribute to fostering inclusive design practices by offering insights into how different cultural groups perceive and interact with design artifacts, thereby helping designers in creating products that resonate universally while respecting cultural specificities.
    \item \textbf{Old Ideas}: LLMs, characterized by their capacity to process vast amounts of data, play an important role in uncovering historical and contemporary design concepts.LLMs can analyze \cite{tai2024examination} archival texts, design critiques, and historical documents to identify seminal ideas and trends across different epochs. By discerning recurring motifs, stylistic evolutions, and influential design paradigms, LLMs facilitate a nuanced understanding of the lineage and evolution of design principles. Furthermore, through comparative analysis and pattern recognition, LLMs enable designers and scholars to trace the interconnectedness of ideas, discerning how past innovations inform present-day practices. This capability not only enriches the discourse surrounding design theory and history but also inspires innovative approaches by juxtaposing established concepts with contemporary sensibilities. 
    % Thus, LLMs serve as invaluable tools in the perpetual exploration and reinterpretation of design heritage, fostering a dynamic dialogue between tradition and innovation in the field of design.
    \item \textbf{Competition}: To discover competitive design ideas, LLMs can employ various systematic approaches. Firstly, they can conduct extensive research and analysis of current trends and user preferences within the specific domain of interest. This involves reviewing existing designs and identifying gaps or areas for improvement. Additionally, LLMs can utilize advanced algorithms to generate and evaluate a wide range of design concepts \cite{ccelen2024design} based on predefined criteria such as functionality, aesthetics, and usability. Collaborative platforms and crowdsourcing methods also offer opportunities for LLMs to gather diverse perspectives and innovative ideas from a global community of designers and stakeholders. 
\end{enumerate}

\subsection{Crafting Dialogue with Designers}
Dialogue holds significant importance within the discipline of design \cite{bjorkman2004design}. It functions not merely as a means of communication but as a pivotal mechanism for exchanging diverse viewpoints, aligning goals, and refining conceptual frameworks. In design practice, effective dialogue nurtures a collaborative environment wherein stakeholders can harmonize their expertise and creative visions, thereby enhancing the coherence and efficacy of design processes and outcomes.

\begin{enumerate}
    \item \textbf{Clarifying Ambiguities}: LLMs, leveraging vast datasets and sophisticated algorithms, offer a unique capability to generate diverse and innovative ideas that are required for clarifying ambiguities. Research indicates that LLMs can facilitate creativity by providing users with a broad spectrum of suggestions \cite{xu2024jamplate,shaer2024ai}, thereby overcoming cognitive biases and expanding the scope of possible solutions. Additionally, the interactive nature of LLMs enables iterative refinement of ideas, allowing users to explore and develop concepts more deeply. As such, while LLMs serve as a valuable tool for stimulating creativity and resolving ambiguities, it remains the responsibility of the designer to consider all possibilities and determine the subsequent steps in the design process.

    \item \textbf{Moving across ethical dilemmas}: LLMs can significantly aid in navigating ethical dilemmas by providing comprehensive analyses of complex situations \cite{shaer2024ai}. They can process vast amounts of information and offer diverse perspectives, which are essential for understanding multifaceted ethical issues. Additionally, LLMs can generate potential solutions and predict their consequences, enabling decision-makers to evaluate the ethical implications of various courses of action. By facilitating access to a broad array of ethical theories and principles, LLMs can help ensure that decisions are well-informed and balanced. Furthermore, the use of LLMs can mitigate cognitive biases and provide a more objective foundation for ethical deliberations. As with clarifying ambiguities, it is for the designer to make the final decision.  
    % Thus, integrating LLMs into ethical decision-making processes holds the promise of enhancing the rigor and depth of ethical analyses in various professional and academic contexts.
    \item \textbf{Overcoming Designers' block}: Writer's block \cite{rose2009writer}, characterized by a temporary inability to generate new ideas or produce written content, is a significant challenge to writers across disciplines. LLMs offer promising potential in assisting writers (and therefore designers) by providing expansive vocabularies, syntactic diversity, contextually relevant suggestions and ideas. Through their ability to generate coherent and contextually appropriate text based on prompts or partial sentences, LLMs can serve as valuable tools for overcoming the mental barriers associated with writer's block. By facilitating idea generation and enhancing productivity, LLMs contribute to alleviating the cognitive burdens that impede design progress. 
\end{enumerate}

\subsection{Design Evaluation}

Evaluation constitutes an indispensable stage within the design process, playing a pivotal role in validating the functionality, efficiency, and user-friendliness of a proposed design solution. Through systematic assessment and feedback collection, designers can discern the extent to which their creations meet predetermined objectives and user needs. 
% This iterative process not only aids in identifying strengths and weaknesses but also facilitates refinement and enhancement of the design prior to implementation. 

\begin{enumerate}
    \item \textbf{Engaging directly with Design} 
    LLMs can capture a variety of data. This capability spans a spectrum of formats ranging from Scalable Vector Graphics (SVG) \cite{xu2024exploring}, Comma-Separated Values (CSV) \cite{vazquez2024llms}, among numerous others. The proficiency of LLMs in comprehending these formats necessitates a multifaceted approach involving parsing, decoding, and understanding the structural intricacies inherent to each file type. This enables the LLMs to conduct a comprehensive critique of the design in its original raw format.
    \item \textbf{A full critique of the design choices} 
    % \cite{alabood2023systematic,luther2015structuring,xu2011crowdsourcing}
    LLMs can systematically analyze and interpret vast amounts of textual data, including design documentation, user feedback, and relevant literature \cite{alabood2023systematic,luther2015structuring}. This enables them to identify underlying assumptions, assess the coherence and alignment of design decisions with stated objectives, and highlight potential inconsistencies or areas for improvement. Furthermore, LLMs can provide insights by cross-referencing contemporary design principles and best practices, thereby offering a nuanced evaluation that encompasses both historical context and emerging trends. Consequently, the integration of LLMs into the critique process not only augments the rigor and depth of the analysis but also enhances the overall quality of design outcomes through data-driven and contextually informed feedback.
    
    \item \textbf{Comparative Analysis with an existing product}
    The application of LLMs in the realm of design analysis offers significant potential for enhancing comparative evaluations of diverse design solutions. LLMs can efficiently analyze and synthesize design specifications, user reviews, technical documentation, and scholarly articles. By employing sophisticated algorithms to identify patterns, extract pertinent information, and generate comprehensive summaries \cite{chew2023llm}, LLMs facilitate a more systematic and thorough comparison of design alternatives. This capability not only streamlines the analysis process but also mitigates the potential for human bias and error, thereby ensuring a more objective and reliable assessment. Furthermore, LLMs can assist in predicting the performance and user acceptance of different designs by integrating insights from historical data and contemporary trends. 
    % Consequently, the incorporation of LLMs into design analysis processes holds promise for significantly advancing the precision and efficacy of comparative design evaluations.
    \item \textbf{Finding out Dark Patterns}
    LLMs possess significant potential to uncover dark patterns  \cite{gray2018dark}, which are manipulative design tactics employed by websites and applications to deceive users into actions they might not otherwise take. By leveraging their advanced natural language processing capabilities, LLMs can systematically analyze user interface descriptions, terms of service, privacy policies, and user reviews. This analysis can identify subtle linguistic cues and patterns indicative of deceptive practices. Furthermore, LLMs can be finetuned to recognize and classify these patterns, thereby providing insights into their prevalence and variations across different platforms. Such automated and scalable analysis can assist regulatory bodies, consumer protection organizations, and researchers in identifying and mitigating the impact of dark patterns, thereby promoting transparency and user autonomy in digital environments.
\end{enumerate}

\section{Discussion}
The proposed framework should be regarded as a dynamic and adaptive guideline rather than a fixed or rigid design space. This adaptability is necessitated by the continuous advancement of LLMs and the emergence of novel datasets, both of which introduce new capabilities that must be systematically integrated into the design process. As these models evolve, their enhanced functionalities and limitations necessitate ongoing reassessment and refinement of the framework to ensure its continued relevance and effectiveness. Consequently, rather than prescribing a static set of principles, the framework serves as a flexible structure that accommodates technological progress and empirical insights. This adaptability enables designers to incorporate state-of-the-art developments in LLMs, thereby optimizing the framework's applicability across different contexts. 
% By maintaining an iterative and responsive approach, the framework aligns with the rapidly evolving landscape of artificial intelligence and supports the development of robust, context-aware, and ethically grounded applications.
\newpage
\bibliography{references}
\end{document}